\documentclass[leqno,twoside]{article}
\usepackage{amssymb,amsmath,latexsym, theorem}

\def\dj{d\kern-0.4em\char"16\kern-0.1em}
\def\Dj{\mbox{\raise0.3ex\hbox{-}\kern-0.4em D}}

\newtheorem{definition}{Definition}[section]
\newtheorem{theorem}{Theorem}[section]

\newtheorem{corollary}{Colorallary}[section]

\title{} \author{} \date{}
\usepackage[dvipdfm, pdfstartview=FitH]{hyperref}
\begin{document}
\thispagestyle{empty}
\pagestyle{myheadings}
%\noindent\rule[2mm]{155mm}{.5mm}

\begin{center}
\noindent{\huge\bf $S$-Program Calculus\\}
\vspace{3mm}
\noindent{\bf ALEKSANDAR KUPUSINAC and DU\v{S}AN MALBA\v{S}KI\footnote{Authors' address: A. Kupusinac and D. Malba\v{s}ki, University of Novi Sad, Faculty of Technical Sciences, Trg Dositeja Obradovi\'{c}a 6, 21000 Novi Sad, Serbia. \{sasak, malbaski\}@uns.ac.rs}}
\end{center}
\vspace{8mm}

\begin{abstract}
This paper presents a special subset of the first-order predicate logic named $S$-program calculus (briefly $S$-calculus). The $S$-calculus is a calculus consisting of so-called $S$-formulas that are defined over the abstract state space of a virtual machine. We show that $S$-formulas are a highly general tool for analyzing program semantics inasmuch as Hoare triplets of total and partial correctness are not more than two $S$-formulas. Moreover, all the rules of Hoare logic can be derived using $S$-formulas and axioms/theorems of first-order predicate calculus. The $S$-calculus is a powerful mechanism for proving program correctness as well as for building additional proving tools using theorems of the predicate logic. Every proof is based on deriving the validity of some $S$-formula, so the procedure may be automated using automatic theorem provers (we will use {\tt Coq} in this paper). As an example of the use of $S$-calculus, we will prove the four basic properties of Dijsktra's operator $wp$. The proofs given by Dijkstra are not completely formalized and we will show that a full formalization can be achieved using $S$-calculus. Finally, we add one more theorem to the above-mentioned four, namely the law of negation.\\


\end{abstract}

\section{Introduction}

The key motivation for this research is the idea that programs may be treated as predicates and/or Boolean expressions \cite{28}\cite{24}\cite{19}\cite{32}\cite{23}\cite{20}\cite{21}. The connection between Floyd-Hoare logic \cite{15}\cite{22} and predicate logic is outlined in the papers of Cook \cite{10} and Blass and Gurevich \cite{7}, where they use it to analyze the completeness of Hoare logic \cite{2}. Blass and Gurevich consider the possibility of incorporating first-order predicate logic into Hoare logic, but they conclude that it would significantly increase the complexity of the latter. In our opinion, it is not necessarily the case: it is possible to generalize the ideas of Hoare logic on the abstract state space and simultaneously simplify proofs, if the interpretation domain is strictly separated from the domain of the abstract state space. Back, Akademi and von Wright \cite{3} have developed the idea of a special program calculus called {\it refinement calculus}, which was meant to combine Hoare's ideas with predicate logic. They solved the problem of indeterminism in the total/partial correctness formulas by introducing additional formulas of angelical and demonical correctness \cite{4}, but at the price of increasing the complexity of refinement calculus. Our idea is to develop a program calculus that associates Hoare logic with first-order predicate logic and clearly separates the interpretation domain from the abstract state domain (similarly to \cite{27}). Secondly, it must not have any problems with indeterminism. Finally, it must treat total/partial correctness directly, i.e. without any requirement for additional concepts and formulas.

In this paper, we will present the development of $S$-program calculus (briefly $S$-calculus), which represents a mathematical tool for the program semantics analysis \cite{26}. Generality of the $S$-calculus stems from the fact that it is built around so-called $S$-formulas that are defined on the abstract state space and not on any of its interpretations, which was the reason for naming it "$S$-calculus", after the word "state". Simultaneously with the development of $S$-calculus, we will discuss the following six issues:

\begin{itemize}
\item[1.)] The $S$-calculus uses an abstract state space and is a general tool for describing program semantics.
\item[2.)] Hoare's formulas of total and partial correctness are no more than two particular $S$-formulas.
\item[3.)] The $S$-calculus is based on the axioms and theorems of first-order predicate logic. The assignment statement and standard syntax units (such as {\it if-then}, {\it if-then-else}, {\it while} etc.) are defined using $S$-formulas so there is no need for special axioms and rules, as in Hoare logic.
\item[4.)] Variable declaration is also described using appropriate $S$-formulas.
\item[5.)] The general rules of Hoare logic are theorems in $S$-calculus and can be derived using axioms and theorems of first-order predicate logic.
\item[6.)] Proofs in the $S$-calculus are simple since they rely only on the results of first-order predicate logic. Moreover, they lend themselves to automation using standard theorem provers, making it easier to introduce new rules and theorems.
\end{itemize}

The axiomatic system of $S$-calculus consists of the axioms of first-order predicate calculus. Each theorem in the predicate calculus is also a theorem in the $S$-calculus and vice versa. In Section \ref{sec:second}, we will present the basic components, the axioms and several theorems in the $S$-calculus.

Hoare logic incorporates the formulas of total and partial correctness, the assignment axiom and numerous rules \cite{1}\cite{17}. The formulas of total and partial correctness are customarily denoted respectively by $\{P\}S\{Q\}$ and $P\{S\}Q$ and their meaning is given in a descriptive form. Instead of this, the $S$-calculus introduces strict mathematical notation for both formulas treating them as two special $S$-formulas. This will be discussed in Section \ref{sec:second}.

In Section \ref{sec:third}, we will show that the general rules of Hoare logic are theorems in the $S$-calculus that can be derived using solely the axioms and theorems of predicate logic. It follows that Hoare logic is a special case of the $S$-calculus and consequently a special case of first-order predicate logic.

Hoare's assignment axiom and the rules for special syntax units are not needed in the $S$-calculus. They are considered as special $S$-formulas or, more precisely, as $S$-relations, where the term "$S$-relation" refers to a binary relation on the abstract state space. For example, the assignment statement $a:=e;$ is an interpretation of the appropriate $S$-relation $S_{a:=e;}$ introduced by definition. In Section \ref{sec:fourth}, we will show the definitions of $S$-relations, the interpretations of which are the statements no-operation, assignment, {\it if-then-else}, {\it if-then}, {\it while} and the sequence. In addition, we will introduce a special $S$-formula whose interpretation is variable declaration. Owing to its generality, the $S$-calculus allows variable definition to be considered as a special syntax unit, which is a problem in theories that use interpreted state space \cite{8}\cite{12}\cite{16}\cite{17}\cite{30}. This opportunity is especially important for programming languages in which the declaration is an ordinary statement (as in Java), because it makes automated correctness proofs possible \cite{27}.

Dijkstra has formulated the four basic theorems concerning the weakest precondition $wp$: the law of the excluded miracle, the law of monotonicity, the law of conjunction and the law of disjunction. From the mathematical point of view, these proofs are not strictly formal \cite{14}\cite{18}. In Section \ref{sec:fifth}, we will prove these theorems in a strictly formal way. In addition, we will prove the fifth basic law, namely the law of negation. Finally, we will provide the formal proof of Dijkstra's theorem on total correctness.

The aim of this paper is not to lessen the importance of Hoare logic. On the contrary, we try to generalize the basic ideas by raising its domain to the level of abstract state space. The $S$-calculus is supposed to serve as a mathematical bridge between Hoare logic and the formalism of classical predicate calculus. Connecting Hoare's ideas with predicate logic is of significant importance. In such connection Hoare logic is an appropriate mechanism for describing program syntax, while in its background predicate logic stays with its powerful mathematical proving tools. Accordingly, proving program correctness \cite{13}\cite{5}, as well as building new theorems in the $S$-calculus conforms to the validity proofs of appropriate $S$-formulas. Based on that, we may conclude that for proving program correctness and new theorems we need rather uncomplicated mathematical tools such as axioms, theorems and proving procedures of first-order predicate logic \cite{9}\cite{11}\cite{25}\cite{29}. Moreover, $S$-calculus lends itself to automation, i.e. the above-mentioned proofs can be automated by using theorem provers. We will demonstrate those possibilities using the prover {\tt Coq} \cite{6}\cite{31}.

\section{The Basic Components}\label{sec:second}

The basic components of the $S$-calculus are:

\begin{itemize}
\item[1.)] The set of abstract states (abstract state space) $A$,
\item[2.)] State variables ($S$-variables) $x, y, z , \dots$,
\item[3.)] State constants ($S$-constants) $s_1, s_2, s_3, \dots$,
\item[4.)] $S$-predicates $P, Q, R, \dots$,
\item[5.)] $S$-relations $S_1, S_2, S_3, \dots$,
\item[6.)] $S$-formulas $F_1, F_2, F_3, \dots$,
\item[7.)] Program variables $a, b, c, \dots$,
\item[8.)] Program constants $c_1, c_2, c_3, \dots$,
\item[9.)] The set of logical operations $\{\lnot, \land, \lor, \Rightarrow, \Leftrightarrow\}$, where $\lnot$ is negation, $\land$ is conjunction, $\lor$ is disjunction, $\Rightarrow$ is implication and $\Leftrightarrow$ is equivalence,
\item[10.)] Set of logical constants $\{\top, \bot\}$ where $\top$ represents true and $\bot$ represents false,
\item[11.)] Brackets $( \ )$ and $[ \ ]$ for changing the priority of operations.
\end{itemize}
	
Each $S$-constant describes an abstract state of the virtual machine. The set $A$ of abstract states is a set of all $S$-constants. $S$-predicates are logical functions over the abstract state space, i.e. $P: A \rightarrow \{\top, \bot\}$. Also, we need two special $S$-predicates $\tau$ and $\phi$ defined by\\

\noindent $(TAU) \hspace{1cm} \forall x\in A, \ \tau(x)=\top $,\\
\noindent $(PHI) \hspace{1cm} \forall x\in A, \ \phi(x)=\bot $.\\

$S$-relations are relations on the abstract state space, i.e. $S\subseteq A\times A$. In other words, $S$-relations are logical functions on the set $A\times A$, i.e. $S: A\times A\rightarrow\{\top, \bot\}$.

\begin{definition}
$S$-formulas are obtained in the following way:
\begin{itemize}
\item[a.)] $S$-predicates and $S$-relations are $S$-formulas.
\item[b.)] If $F_1$ and $F_2$ are $S$-formulas then $\lnot F_1, F_1\land F_2, F_1\lor F_2, F_1\Rightarrow F_2, F_1\Leftrightarrow F_2$ are also $S$-formulas.
\item[c.)] Any formula obtained from a.) and b.) in a finite number of steps is an $S$-formula.
\end{itemize}
\end{definition}

Let $\{v_1, v_2, \dots, v_n\} $ be a set of program variables, which take values from sets $D_1, D_2, \dots, D_n$ respectively. Let $A'$ be a subset of $A$, with the cardinality $Card(A')=Card(D_1\times D_2\times \dots \times D_n)$. Interpretation of the set $A$ with respect to the set $\{ v_1, v_2, \dots, v_n \}$ is a bijection that maps any $S$-constant from $A'$ to the appropriate vector of program constants from $D_1, D_2, \dots, D_n$ (usually called state vector). $S$-relation $S(x,y)$ contains ordered pairs $(x,y)$, where $x\in A$ is the initial state and $y\in A$ is the final state. Interpreted restriction of $S$-relation on the set $A'$ is called syntactic unit on program variables $\{ v_1, v_2, \dots, v_n \}$. A syntactic unit may be written in many different ways (program code is one of them), and it can refer to a statement, block, subprogram or program. We will use a term "predicate" for an interpreted restriction of $S$-predicate on the set $A'$, knowing that it is actually a boolean expression over program variables $\{ v_1, v_2, \dots, v_n\}$. This means that we observe two domains: the abstract state domain with $S$-constants, $S$-variables, $S$-predicates and $S$-relations and the interpretation domain with vectors of program constants, program variables, predicates and syntactic units. To simplify, $S$-constant is interpreted as a vector of program constants from the set $D_1, D_2, \dots, D_n$, $S$-predicate is interpreted as a boolean expression, and $S$-relation as a syntactic unit with program variables $\{v_1, v_2, \dots, v_n\}$. Interpretation is denoted by "$:$". For example, $x: a>0 \land b=5$ means that $S$-variable $x$ represents all states in which program variables $a$ and $b$ satisfy $a>0$ and $b=5$.

The symbol $\leftrightarrow$ means "abbreviation". If $\alpha$ is a token and $F$ is an $S$-formula  then $\alpha\leftrightarrow F$ means "$\alpha$ is an abbreviation for $F$". If $F_1$ and $F_2$ are two $S$-formulas with the same form, we say that $F_1$ is syntactically identical to $F_2$, and write $F_1=F_2$. If $F_1$ and $F_2$ have the same meaning but not the same form, they are semantically equivalent, denoted by $F_1\equiv F_2$.

$S$-calculus consists of $S$-formulas and is based solely on axioms and theorems of the first-order predicate logic. It means that, among other things, the formulas $\{P\}S\{Q\}$ and $P\{S\}Q$ are just two $S$-formulas:
\begin{itemize}
\item[a.)] Total correctness formula $(TCF)$:
$$\forall x[P(x)\Rightarrow (\exists y S(x,y) \land \forall z(S(x,z) \Rightarrow Q(z)))] \ .$$
\item[b.)] Partial correctness formula $(PCF)$:
$$ \forall x[(P(x) \land \exists y S(x,y)) \Rightarrow \forall z(S(x,z) \Rightarrow Q(z))] \ .$$
\end{itemize}

When writing $S$-formulas we will obey the usual priority conventions, where the order of priority is: negation $ \lnot $, conjunction $ \land $, disjunction $ \lor $, implication $ \Rightarrow $, equivalence $ \Leftrightarrow $. The priority can be changed by using brackets $( \ )$ and $[ \ ]$.

Firstly, by using the formulas $(TCF)$ and $(PCF)$ we can formally define the total and partial correctness of an $S$-relation with respect to $S$-predicates:

\begin{definition}
$S$-relation $S$ is totally correct with respect to precondition $P$ and postcondition $Q$ if the $S$-formula $\forall x[P(x) \Rightarrow (\exists y S(x,y) \land \forall z(S(x,z) \Rightarrow Q(z)))]$ is valid.
\end{definition}

\begin{definition}
$S$-relation $S$ is partially correct with respect to precondition $P$ and postcondition $Q$ if the $S$-formula $\forall x[(P(x) \land \exists y S(x,y)) \Rightarrow \forall z(S(x,z) \Rightarrow Q(z))]$ is valid.
\end{definition}

Hoare's total correctness formula, denoted by $\{P\}S\{Q\} $, is defined by the statement "if the syntax unit $S$ starts in a state satisfying the predicate $P$, then it terminates in a state satisfying the predicate $Q$" \cite{17}. The connection between this sentence and the formula $(TCF)$ is apparent: if for every state $x$ the $S$-predicate $P$ holds, then the $S$-formula $\forall x \exists y S(x,y) \land \forall x \forall z(S(x,z) \Rightarrow Q(z))$ is true. The state $x$ is then called the initial state. The formula $\forall x \exists y S(x,y)$ means that for every initial state $x$ there exists a state $y$ such that $(x,y) \in S$. The state $y$ is then called the final state. The meaning of the $S$-formula $\forall x \forall z(S(x,z) \Rightarrow Q(z))$ is the following: if for every initial state $x$ and every state $z$ it is true that $(x,z) \in S$, then in the state $z$ the $S$-predicate $Q$ is true.

Hoare's partial correctness formula, denoted by $P\{S\}Q$, is defined by the statement "if the syntax unit $S$ starts in a state satisfying the predicate $P$ and if it terminates then the final state satisfies the predicate $Q$" \cite{17}. In terms of the $S$-calculus, we assert: if for some state $x$ the predicate $P$ holds and if there exists a final state $y$ such that $(x,y) \in S$, then the formula $\forall x \forall z(S(x,z) \Rightarrow Q(z))$ is true.

Concerning the question of indeterminism, $S$-calculus does not require any additional formulas such as angelical or demonical formulas in refinement calculus \cite{3}, because the formulas $(TCF)$ and $(PCF)$ contain $\forall x \forall z(S(x,z) \Rightarrow Q(z))$. Thus, $S$-calculus strictly implements Dijkstra's statement "Eventually I came to regard nondeterminacy as the normal situation, determinacy being reduced to a -– not even very interesting -– special case" \cite{14}.

The $S$-calculus is a special kind of the first-order predicate logic or, more precisely, it is a predicate logic over $S$-formulas. Its axiomatic system consists solely of the predicate logic axioms, provided that formulas $F$, $G$ and $H$ are now $S$-formulas:\\

\noindent $(A_1) \hspace{1cm} F \Rightarrow (G \Rightarrow F) $\\
\noindent $(A_2) \hspace{1cm} (F \Rightarrow (G \Rightarrow H)) \Rightarrow ((F \Rightarrow G) \Rightarrow (F \Rightarrow H)) $\\
\noindent $(A_3) \hspace{1cm} (\lnot F \Rightarrow \lnot G) \Rightarrow (G \Rightarrow F) $\\
\noindent $(A_4) \hspace{1cm} \forall x F(x) \Rightarrow F(t) $ (term $t$ is free for $x$ in $F(t)$)\\
\noindent $(A_5) \hspace{1cm} \forall x(F \Rightarrow G) \Rightarrow (F \Rightarrow \forall x G) $ (variable $x$ is not free in $F$)\\

The $S$-calculus uses the rules of inference from the first-order predicate logic (the symbols $F$ and $G$ stand for $S$-formulas):

\begin{itemize}
\item[a.)] Modus ponens $(MPN)$:
$$\frac {F, F \Rightarrow G}{G}$$
\item[b.)] Generalisation $(GEN):$
$$\frac {F}{\forall x F}$$
\end{itemize}

All theorems i.e. valid formulas in predicate logic are also valid in the $S$-calculus and vice versa. We will briefly cite some well-known theorems of predicate logic that will be needed for further proofs in this paper (again, the symbols $F$, $G$, $H$ and $K$ stand for $S$-formulas and $\tau$ and $\phi$ are defined by $(TAU)$ and $(PHI)$ respectively):\\

\noindent $(T_1) \hspace{1.14cm} \forall x\forall y F\Leftrightarrow\forall y\forall x F$\\
\noindent $(T_2) \hspace{1.14cm} \exists x\forall y F\Rightarrow\forall y\exists x F$\\
\noindent $(T_3) \hspace{1.14cm} \forall x F\Leftrightarrow F$\\
\noindent $(T_4) \hspace{1.14cm} \forall x(F\land G)\Leftrightarrow\forall x F\land\forall x G$\\
\noindent $(T_5) \hspace{1.14cm} \forall x F\lor\forall x G\Rightarrow\forall x(F\lor G)$\\
\noindent $(T_6) \hspace{1.14cm} \lnot\forall x F\Leftrightarrow\exists x\lnot F$\\
\noindent $(T_7) \hspace{1.14cm} \forall x F\Leftrightarrow\forall x(F\Leftrightarrow\tau)$\\
\noindent $(T_8) \hspace{1.14cm} \forall x(\tau\Rightarrow F)\Leftrightarrow\forall x F$\\
\noindent $(T_9) \hspace{1.14cm} \forall x\lnot F\Leftrightarrow\forall x(F\Leftrightarrow\phi)$\\
\noindent $(T_{10}) \hspace{1cm} \forall x(F\Leftrightarrow F\land F)$\\
\noindent $(T_{11}) \hspace{1cm} \forall x(F\Leftrightarrow F\lor G)$\\
\noindent $(T_{12}) \hspace{1cm} \forall x(\lnot F\lor\lnot G)\Leftrightarrow\forall x\lnot(F\land G)$\\
\noindent $(T_{13}) \hspace{1cm} \forall x(\lnot F\land\lnot G)\Leftrightarrow\forall x\lnot(F\lor G)$\\
\noindent $(T_{14}) \hspace{1cm} \forall x(F\Rightarrow G)\Rightarrow(\forall x F\Rightarrow\forall x G)$\\
\noindent $(T_{15}) \hspace{1cm} \forall x(F\Rightarrow G)\Leftrightarrow\forall x(\lnot F\lor G)$\\
\noindent $(T_{16}) \hspace{1cm} \forall x[(F\Rightarrow H)\land(H\Rightarrow G)]\Rightarrow\forall x(F\Rightarrow G)$\\
\noindent $(T_{17}) \hspace{1cm} \forall x[(F\Rightarrow G)\land(H\Rightarrow K)]\Rightarrow\forall x[(F\lor H)\Rightarrow(G\lor K)]$\\
\noindent $(T_{18}) \hspace{1cm} \forall x[(F\Rightarrow G)\land(H\Rightarrow K)]\Rightarrow\forall x[(F\land H)\Rightarrow(G\land K)]$\\
\noindent $(T_{19}) \hspace{1cm} \forall x[(F\Rightarrow G)\land(F\Rightarrow H)]\Leftrightarrow\forall x(F\Rightarrow G\land H)$\\
\noindent $(T_{20}) \hspace{1cm} \forall x[(F\Rightarrow G)\land(F\Rightarrow H)]\Leftrightarrow\forall x(F\Rightarrow G\lor H)$\\
\noindent $(T_{21}) \hspace{1cm} \forall x[(F\Rightarrow H)\land(G\Rightarrow H)]\Leftrightarrow\forall x(F\lor G\Rightarrow H)$\\
\noindent $(T_{22}) \hspace{1cm} \forall x[(F\Rightarrow G)\lor(H\Rightarrow K)]\Rightarrow\forall x[(F\land H)\Rightarrow(G\lor K)]$\\

Program correctness or a new $S$-calculus theorem are proven by proving the validity of an appropriate $S$-formula. This needs a modest mathematical apparatus e.g. the axioms, theorems and proof procedures of the first-order predicate logic. Moreover, it can be automated using various automatic theorem provers.

An important detail is also the fact that the $S$-calculus is based on the abstract set of states $A$. This means that when applying the Hoare logic we do not need an exact description of every abstract state, thus avoiding the use of the program state vector (vector of all program variables). It is known that the use of state vector introduces certain difficulties, since it is not quite clear how to model unknown values of program variables \cite{14}. In addition, the state vector is associated with the specific program, and can not be related to the virtual machine when the program is not active. Subprograms also contribute to the problem because they have their own state vectors. On the other hand, the abstract state space is associated with the virtual machine itself so it is always meaningful, regardless of whether a particular program is active or not. In the $S$-calculus the program state space $A'$ is a subset of the virtual machine abstract state space $A$ ($A'\subseteq A$), and every program is a restriction on $A'$ of the appropriate $S$-relation where $S \subseteq A \times A$.

\section{General Laws of the Hoare logic}\label{sec:third}

In this section we will consider the general laws of Hoare logic \cite{1}\cite{17} such as the laws of consequence, disjunction, conjunction and negation. While the Hoare logic treats these laws as rules, we will treat them as theorems. Some of them will be proven using {\tt Coq} automatic prover.

\begin{theorem}[Laws of Consequence]\label{theorem:conseqence}
The following $S$-formulas are valid:
\begin{itemize}
\item[a.)] $\forall x(P(x)\Rightarrow R(x))\land \{R\}S\{Q\} \ \Rightarrow \ \{P\}S\{Q\}$,
\item[b.)] $\{P\}S\{R\}\land \forall x(R(x)\Rightarrow Q(x)) \ \Rightarrow \ \{P\}S\{Q\}$,
\item[c.)] $\forall x(U(x)\Rightarrow P(x))\land \forall x(Q(x)\Rightarrow V(x))\land \{P\}S\{Q\} \ \Rightarrow \ \{U\}S\{V\}$.
\end{itemize}
\noindent{\bf Proof.}
$ $\phantom{We prove theorems:}
\begin{itemize}
\item[a.)] Since:\\
$\{R\}S\{Q\} \ \leftrightarrow \ \forall x[R(x)\Rightarrow (\exists yS(x,y)\land \forall z(S(x,z)\Rightarrow Q(z)))]$,\\
the left side of the implication can be written as:\\
$\forall x(P(x)\Rightarrow R(x))\land \forall x[R(x)\Rightarrow (\exists yS(x,y)\land \forall z(S(x,z)\Rightarrow Q(z)))]$\\
and by the Theorem $(T_{16})$, we obtain:\\
$\forall x[P(x)\Rightarrow (\exists yS(x,y)\land \forall z(S(x,z)\Rightarrow Q(z)))]$,\\
i.e.\\
$\{P\}S\{Q\}$.\vspace{0.2cm}
\item[b.)] Since:\\
$\{P\}S\{R\} \ \leftrightarrow \ \forall x[P(x)\Rightarrow (\exists yS(x,y)\land \forall z(S(x,z)\Rightarrow R(z)))]$,\\
the left side of the implication can be written as:\\
$\forall x[P(x)\Rightarrow (\exists yS(x,y)\land \forall z(S(x,z)\Rightarrow R(z)))] \ \land \ \forall x(R(x)\Rightarrow Q(x))$\\
and by the Theorem $(T_{16})$, we obtain:\\
$\forall x[P(x)\Rightarrow (\exists yS(x,y)\land \forall z(S(x,z)\Rightarrow Q(z)))]$,\\
i.e.\\
$\{P\}S\{Q\}$.\vspace{0.2cm}
\item[c.)] By the Theorem \ref{theorem:conseqence}$.a.)$, the left side of the implication can be written as:\\
$\forall x(Q(x)\Rightarrow V(x)) \ \land \ \{U\}S\{Q\}$\\
and by the Theorem \ref{theorem:conseqence}$.b.)$, we obtain:\\
$\{U\}S\{V\}$.
\end{itemize}
$\square$
\end{theorem}

The Theorems \ref{theorem:conseqence} also can be proven using automatic prover {\tt Coq} (Appendix \ref{appendix}). Finally, using by the Theorem $(T_3)$, from Theorems \ref{theorem:conseqence} we can obtain the well-known Hoare's rules of consequence \cite{1}\cite{17}:\\
$$\frac{(P\Rightarrow R), \{R\}S\{Q\}}{\{P\}S\{Q\}}, \hspace{7mm} \frac{\{P\}S\{R\}, (R\Rightarrow Q)}{\{P\}S\{Q\}} \hspace{7mm} {\rm and} \hspace{7mm} \frac{(U\Rightarrow P), (Q\Rightarrow V), \{P\}S\{Q\}}{\{U\}S\{V\}}.$$\\

\begin{theorem}[Laws of Conjunction]\label{theorem:conjunction}
The following $S$-formulas are valid:
\begin{itemize}
\item[a.)] $\{P\}S\{Q\}\land \{R\}S\{W\} \ \Rightarrow \ \{P\lor R\}S\{Q\lor W\}$,
\item[b.)] $\{P\}S\{Q\}\land \{R\}S\{W\} \ \Rightarrow \ \{P\land R\}S\{Q\land W\}$.
\end{itemize}
{\bf Proof.}
$ $\phantom{We prove theorems:}
\begin{itemize}
\item[a.)] Since:\\
$\{P\}S\{Q\} \ \leftrightarrow \ \forall x[P(x)\Rightarrow (\exists yS(x,y)\land \forall z(S(x,z)\Rightarrow Q(z)))]$,\\
$\{R\}S\{W\} \ \leftrightarrow \ \forall x[R(x)\Rightarrow (\exists yS(x,y)\land \forall z(S(x,z)\Rightarrow W(z)))]$,\\
by the Theorem $(T_{17})$, the left side of the implication can be written as:\\
$\forall x[(P(x)\lor R(x))\Rightarrow ((\exists yS(x,y)\land \forall z(S(x,z)\Rightarrow Q(z)))\lor(\exists yS(x,y)\land \forall z(S(x,z)\Rightarrow W(z))))]$\\
$\equiv \ \forall x[(P(x)\lor R(x))\Rightarrow \exists y((S(x,y)\land \forall z(S(x,z)\Rightarrow Q(z)))\lor(S(x,y)\land \forall z(S(x,z)\Rightarrow W(z))))]$\\
$\equiv \ \forall x[(P(x)\lor R(x))\Rightarrow \exists yS(x,y)\land \forall z((S(x,z)\Rightarrow Q(z))\lor(S(x,z)\Rightarrow W(z)))]$.\\
Then, by the Theorem $(T_{15})$, we obtain:\\
$\forall x[(P(x)\lor R(x))\Rightarrow \exists yS(x,y)\land \forall z(\lnot S(x,z)\lor Q(z)\lor\lnot S(x,z)\lor W(z))]$\\
$\equiv \ \forall x[(P(x)\lor R(x))\Rightarrow \exists yS(x,y)\land \forall z(\lnot S(x,z)\lor Q(z)\lor W(z))]$\\
and after that, by the Theorem $(T_{15})$, we obtain:\\
$\forall x[(P(x)\lor R(x))\Rightarrow \exists yS(x,y)\land \forall z(S(x,z)\Rightarrow (Q(z)\lor W(z)))]$,\\
i.e.\\
$\{P\lor R\}S\{Q\lor W\}$.\vspace{0.2cm}
\item[b.)] By the Theorem $(T_{18})$, the left side of the implication can be written as:\\
$\forall x[(P(x)\land R(x))\Rightarrow ((\exists yS(x,y)\land \forall z(S(x,z)\Rightarrow Q(z)))\land(\exists yS(x,y)\land \forall z(S(x,z)\Rightarrow W(z))))]$\\
$\equiv \ \forall x[(P(x)\land R(x))\Rightarrow (\exists yS(x,y)\land \forall z(S(x,z)\Rightarrow Q(z))\land \forall z(S(x,z)\Rightarrow W(z)))]$\\
$\equiv \ \forall x[(P(x)\land R(x))\Rightarrow \exists yS(x,y)\land \forall z((S(x,z)\Rightarrow Q(z))\land (S(x,z)\Rightarrow W(z)))]$\\
and by the Theorem $(T_{19})$, we obtain:\\
$\forall x[(P(x)\land R(x))\Rightarrow \exists yS(x,y)\land \forall z(S(x,z)\Rightarrow (Q(z)\land W(z)))]$,\\
i.e.\\
$\{P\land R\}S\{Q\land W\}$.
\end{itemize}
$\square$
\end{theorem}

\begin{corollary}[Law of Resolution]\label{corollary:resolution}
The following $S$-formula is valid:

\vspace{0.215cm}\hspace{0.15cm} $\{P\}S\{Q\}\land \{\lnot P\}S\{W\} \ \Rightarrow \ \{\tau\}S\{Q\lor W\}$.\\

{\bf Proof.}
$ $\phantom{We prove corollary:}
\begin{itemize}
\item[ ] If we substitute $R$ with $\lnot P$ in the Theorem \ref{theorem:conjunction}$.a.)$ we obtain:\\
$\{P\}S\{Q\}\land \{\lnot P\}S\{W\} \ \Rightarrow \ \{P\lor\lnot P\}S\{Q\lor W\}$,\\
i.e.\\
$\{P\}S\{Q\}\land \{\lnot P\}S\{W\} \ \Rightarrow \ \{\tau\}S\{Q\lor W\}$.
\end{itemize}
$\square$
\end{corollary}

\begin{theorem}[Laws of Disjunction]\label{theorem:disjunction}
The following $S$-formula is valid:

\vspace{0.215cm}\hspace{0.15cm} $\{P\}S\{Q\}\lor \{R\}S\{W\} \ \Rightarrow \ \{P\land R\}S\{Q\lor W\}$.\\

{\bf Proof.}
$ $\phantom{We prove theorem:}
\begin{itemize}
\item[ ] Since:\\
$\{P\}S\{Q\} \ \leftrightarrow \ \forall x[P(x)\Rightarrow (\exists yS(x,y)\land \forall z(S(x,z)\Rightarrow Q(z)))]$,\\
$\{R\}S\{W\} \ \leftrightarrow \ \forall x[R(x)\Rightarrow (\exists yS(x,y)\land \forall z(S(x,z)\Rightarrow W(z)))]$,\\
by the Theorem $(T_{22})$, the left side of implication can be written as:\\
$\forall x[(P(x)\land R(x))\Rightarrow (\exists yS(x,y)\land \forall z(S(x,z)\Rightarrow Q(z)))\lor(\exists yS(x,y)\land \forall z(S(x,z)\Rightarrow W(z)))]$\\
$\equiv \ \forall x[(P(x)\land R(x))\Rightarrow \exists yS(x,y)\land \forall z((S(x,z)\Rightarrow Q(z))\lor(S(x,z)\Rightarrow W(z)))]$\\
and by the Theorem $(T_{20})$, we obtain:\\
$\forall x[(P(x)\land R(x))\Rightarrow \exists yS(x,y)\land \forall z(S(x,z)\Rightarrow (Q(z)\lor W(z)))]$,\\
i.e.\\
$\{P\land R\}S\{Q\lor W\}$.
\end{itemize}
$\square$
\end{theorem}

The proves of Theorems \ref{theorem:conjunction}$.a)$, \ref{theorem:conjunction}$.b)$ and \ref{theorem:disjunction} in {\tt Coq} are given in Appendix \ref{appendix}.

\begin{theorem}[Laws of Conjunction and Disjunction]\label{theorem:conjunction_disjunction}
The following $S$-formulas are valid:
\begin{itemize}
\item[a.)] $\{P\lor R\}S\{Q\} \ \Leftrightarrow \ \{P\}S\{Q\}\land\{R\}S\{Q\}$,
\item[b.)] $\{P\}S\{Q\land R\} \ \Leftrightarrow \ \{P\}S\{Q\}\land\{P\}S\{R\}$,
\item[c.)] $\{P\lor U\}S\{Q\land W\} \ \Leftrightarrow \ \{P\}S\{Q\}\land\{U\}S\{W\}\land\{P\}S\{W\}\land\{U\}S\{Q\}$,
\item[d.)] $\{P\}S\{Q\}\lor \{P\}S\{W\} \ \Rightarrow \ \{P\}S\{Q\lor W\}$.
\end{itemize}
{\bf Proof.}
$ $\phantom{We prove theorems:}
\begin{itemize}
\item[a.)] The left side of the eqivalence can be written as:\\
$\forall x[(P(x)\lor R(x))\Rightarrow \exists yS(x,y)\land\forall z(S(x,z)\Rightarrow Q(z))]$\\
and by the Theorem $(T_{21})$, we obtain:\\
$\forall x[(P(x)\Rightarrow \exists yS(x,y)\land \forall z(S(x,z)\Rightarrow Q(z))) \land (R(x)\Rightarrow \exists yS(x,y)\land \forall z(S(x,z)\Rightarrow Q(z)))]$,\\
i.e.\\
$\{P\}S\{Q\}\land\{R\}S\{Q\}$.\vspace{0.2cm}
\item[b.)] The left side of the eqivalence can be written as:\\
$\forall x[P(x)\Rightarrow \exists yS(x,y)\land \forall z(S(x,z)\Rightarrow (Q(z)\land R(z)))]$.\\
Then, by the Theorem $(T_{19})$, we obtain:\\
$\forall x[P(x)\Rightarrow \exists yS(x,y)\land \forall z(S(x,z)\Rightarrow Q(z)) \land \forall z(S(x,z)\Rightarrow R(z))]$\\
and after that, by the Theorem $(T_{10})$, we obtain:\\
$\forall x[P(x)\Rightarrow \exists yS(x,y)\land \forall z(S(x,z)\Rightarrow Q(z)) \land \exists yS(x,y) \land \forall z(S(x,z)\Rightarrow R(z))]$\\
and finally, by the Theorem $(T_{19})$, we obtain:\\
$\forall x[P(x)\Rightarrow \exists yS(x,y)\land \forall z(S(x,z)\Rightarrow Q(z))] \land \forall x[P(x)\Rightarrow \exists yS(x,y) \land \forall z(S(x,z)\Rightarrow R(z))]$,\\
i.e.\\
$\{P\}S\{Q\}\land\{P\}S\{R\}$.\vspace{0.2cm}
\item[c.)] By the Theorem \ref{theorem:conjunction_disjunction}$.a.)$, from the left side of the equivalence we obtain:\\
$\{P\lor U\}S\{Q\land W\} \Leftrightarrow \{P\}S\{Q\land W\}\land \{U\}S\{Q\land W\}$\\
and by the Theorem \ref{theorem:conjunction_disjunction}$.b.)$, we obtain:\\
$\{P\}S\{Q\land W\}\land\{U\}S\{Q\land W\}  \Leftrightarrow \{P\}S\{Q\}\land\{U\}S\{W\}\land\{P\}S\{W\}\land\{U\}S\{Q\}$.\vspace{0.2cm}
\item[d.)] If we substitute $R$ with $P$ in the Theorem \ref{theorem:disjunction} we obtain:\\
$\{P\}S\{Q\}\lor\{P\}S\{W\} \Rightarrow \{P\land P\}S\{Q\lor W\}$\\
and by the Theorem $(T_{10})$, we obtain:\\
$\{P\}S\{Q\}\lor\{P\}S\{W\} \Rightarrow \{P\}S\{Q\lor W\}$.
\end{itemize}
$\square$
\end{theorem}

\begin{theorem}[General Law of the Excluded Miracle]\label{theorem:excluded_miracle}
The following $S$-formula is valid:

\vspace{0.215cm}\hspace{0.15cm} $\{P\}S\{\phi\}\Leftrightarrow(P\Leftrightarrow\phi)$, i.e. $\{P\}S\{\phi\}\Leftrightarrow\lnot P$.\\

{\bf Proof.}
$ $\phantom{We prove theorem:}
\begin{itemize}
\item[ ] The left side of the equivalence can be written as:\\
$\forall x[P(x)\Rightarrow (\exists yS(x,y)\land\forall z(S(x,z)\Rightarrow\phi(z)))]$.\\
Since the $S$-formula $\forall x\forall z(S(x,z)\Rightarrow\phi(z))$ is valid if $\forall x\forall z\lnot S(x,z)$ is valid, we obtain:\\
$\forall x[P(x)\Rightarrow (\exists yS(x,y)\land\forall z\lnot S(x,z))]$\\
$\equiv \ \forall x[P(x)\Rightarrow\phi(x)]$\\
and by the Theorem $(T_{9})$, we obtain:\\
$\forall x\lnot P(x)$,\\
i.e.\\
$\lnot P$.
\end{itemize}
$\square$
\end{theorem}

\begin{theorem}[Laws of Negation]\label{theorem:negation}
The following $S$-formulas are valid:
\begin{itemize}
\item[a.)] $\{P\}S\{Q\}\land\{R\}S\{\lnot Q\} \ \Rightarrow \ \lnot(P\land R)$,
\item[b.)] $\{P\}S\{Q\}\land\{P\}S\{\lnot Q\} \ \Leftrightarrow \ \forall x\lnot P(x)$,
\item[c.)] $[\{P\}S\{\lnot Q\}\Rightarrow\lnot\{P\}S\{Q\}] \ \Leftrightarrow \ \exists xP(x)$,
\item[d.)] $\{P\}S\{Q\}\land\{\lnot P\}S\{Q\} \ \Leftrightarrow \ \forall x\exists yS(x,y)\land \forall x\forall z(S(x,z)\Rightarrow Q(x))$,
\item[e.)] $\exists x\exists zS(x,z)\land\lnot Q(z) \ \Rightarrow \ [\{\lnot P\}S\{Q\}\Rightarrow\lnot \{P\}S\{Q\}]$.
\end{itemize}
{\bf Proof.}
$ $\phantom{We prove theorems:}
\begin{itemize}
\item[a.)] If we substitute $W$ with $\lnot Q$ in the Theorem \ref{theorem:conjunction}$.b.)$, we obtain:\\
$\{P\}S\{Q\}\land\{R\}S\{\lnot Q\} \Rightarrow \{P\land R\}S\{Q\land\lnot Q\}$\\
$\equiv \ \{P\}S\{Q\}\land\{R\}S\{\lnot Q\} \Rightarrow \{P\land R\}S\{\phi\}$\\
and by the Theorem \ref{theorem:excluded_miracle}, we obtain:\\
$\{P\}S\{Q\}\land\{R\}S\{\lnot Q\} \Rightarrow \lnot(P\land R)$.\vspace{0.2cm}
\item[b.)] If we substitute $R$ with $\lnot Q$ in the Theorem \ref{theorem:conjunction_disjunction}$.b.)$, we obtain:\\
$\{P\}S\{Q\}\land\{P\}S\{\lnot Q\} \Leftrightarrow \{P\}S\{Q\land\lnot Q\}$\\
$\equiv \ \{P\}S\{Q\}\land\{P\}S\{\lnot Q\} \Leftrightarrow \{P\}S\{\phi\}$\\
and by the Theorem \ref{theorem:excluded_miracle}, we obtain:\\
$\{P\}S\{Q\}\land\{P\}S\{\lnot Q\} \Leftrightarrow \lnot P$,\\
i.e.\\
$\{P\}S\{Q\}\land\{P\}S\{\lnot Q\} \Leftrightarrow \forall x\lnot P(x)$.\vspace{0.2cm}
\item[c.)] By the Theorem $(T_{15})$, the left side of the equivalence become:\\
$\lnot\{P\}S\{\lnot Q\} \lor \lnot\{P\}S\{Q\}$\\
and subsequently, by the Theorem $(T_{12})$, we obtain:\\
$\lnot[\{P\}S\{\lnot Q\} \land \{P\}S\{Q\}]$.\\
Then, by the Theorem \ref{theorem:negation}$.b.)$, we obtain:\\
$\lnot[\forall x\lnot P(x)]$\\
and after that, by the Theorem $(T_{6})$, we obtain:\\
$\exists xP(x)$.\vspace{0.2cm}
\item[d.)] If we substitute $R$ with $\lnot P$ in the Theorem \ref{theorem:conjunction_disjunction}$.a.)$, we obtain:\\
$\{P\lor\lnot P\}S\{Q\}$\\
$\equiv \ \{\tau\}S\{Q\}$.\\
Since:\\
$\{\tau\}S\{Q\} \ \leftrightarrow \ \forall x[\tau(x)\Rightarrow\exists yS(x,y)\land\forall z(S(x,z)\Rightarrow Q(z))]$,\\
by the Theorem $(T_{8})$, we obtain:\\
$\forall x[\exists yS(x,y)\land\forall z(S(x,z)\Rightarrow Q(z))]$.\vspace{0.2cm}
\item[e.)] By the Theorem $(T_{15})$, the right side of the equivalence can be written as:\\
$\lnot\{\lnot P\}S\{Q\}\lor\lnot\{P\}S\{Q\}$\\
and by the Theorem $(T_{12})$, we obtain:\\
$\lnot[\{\lnot P\}S\{Q\}\land\{P\}S\{Q\}]$.\\
Then, by the Theorem \ref{theorem:negation}$.d.)$, we obtain:\\
$\lnot\forall x[\exists yS(x,y)\land\forall z(S(x,z)\Rightarrow Q(z))]$\\
$\equiv \ \exists x\lnot[\exists yS(x,y)\land\forall z(S(x,z)\Rightarrow Q(z))]$.\\
By the Theorem $(T_{13})$, we obtain:\\
$\exists x[\lnot\exists yS(x,y)\lor\lnot\forall z(S(x,z)\Rightarrow Q(z))]$\\
$\equiv \ \exists x[\forall y\lnot S(x,y)\lor\exists z\lnot(S(x,z)\Rightarrow Q(z))]$\\
and by the Theorem $(T_{15})$, we obtain:\\
$\exists x[\forall y\lnot S(x,y)\lor\exists z\lnot(\lnot S(x,z)\lor Q(z))]$.\\
After that, by the Theorem $(T_{12})$, we obtain:\\
$\exists x[\forall y\lnot S(x,y)\lor\exists z(S(x,z)\land \lnot Q(z))]$\\
$\equiv \ \exists x\forall y\lnot S(x,y)\lor\exists x\exists z(S(x,z)\land \lnot Q(z))$\\
and finally, by the Theorem $(T_{11})$, we obtain:\\
$\exists x\exists z(S(x,z)\land\lnot Q(z)) \Rightarrow \exists x\exists z(S(x,z)\land\lnot Q(z))\lor\exists x\forall y\lnot S(x,y)$.\\
\end{itemize}
$\square$
\end{theorem}

The proves of Theorems \ref{theorem:excluded_miracle}, \ref{theorem:negation}$.a)$ and \ref{theorem:negation}$.b)$ in {\tt Coq} are given in Appendix \ref{appendix}.

\begin{corollary}\label{corollary:negation1}
The following $S$-formula is valid:

\vspace{0.215cm}\hspace{0.15cm} $\{P\}S\{Q\}\land \{P\}S\{\lnot Q\} \ \Leftrightarrow \ (P\Leftrightarrow \phi)$.\\

{\bf Proof.}
$ $\phantom{We prove corollary:}
\begin{itemize}
\item[ ] From the Theorem \ref{theorem:negation}$.b.)$ we obtain:\\
$\forall x\lnot P(x)$,\\
i.e.\\
$(P\Leftrightarrow \phi)$.\\
\end{itemize}
$\square$
\end{corollary}

\begin{corollary}\label{corollary:negation2}
The following $S$-formula is valid:

\vspace{0.215cm}\hspace{0.15cm} $[\{P\}S\{\lnot Q\}\Rightarrow \lnot\{P\}S\{Q\}] \ \Leftrightarrow \ \lnot(P\Leftrightarrow\phi)$.\\

{\bf Proof.}
$ $\phantom{We prove corollary:}
\begin{itemize}
\item[ ] From Theorem \ref{theorem:negation}$.c.)$ we obtain:\\
$\exists xP(x)$\\
$\equiv \ \lnot(\forall x\lnot P(x))$,\\
i.e.\\
$\lnot(P\Leftrightarrow\phi)$.
\end{itemize}
$\square$
\end{corollary}

\section{Special {\it S}-Relations}\label{sec:fourth}

In Hoare logic, the so-called special syntax units such as {\it if-then}, {\it if-then-else}, {\it while} etc. are introduced by rules, whilst the assignment is defined through an axiom \cite{1}\cite{17}. In the $S$-calculus, things are different, since all special syntax units are treated as particular $S$-formulas. In other words, for every syntax unit we define an appropriate $S$-relation that is a subset of the set $A\times A$ where $A$ is the abstract state space. We will discuss the meaning of this on the example of assignment. Let $a$ be a program variable of the type $integer$. The syntax unit $a:=5;$ is an interpretation of the $S$-relation $S_{a:=5;}$, which transfers the virtual machine from the state $x$ to the state $y$. The state $x$ is interpreted as a state in which the variable $a$ has some value from the its domain $D_{integer}$ (i.e. $a\in D_{integer}$), and the state $y$ is interpreted as a state in which the variable $a$ has the value $5$ (i.e. $a=5$). Accordingly, we define the $S$-relation $S_{a:=5;}$ as a set of ordered pairs $(x,y)$, $x,y\in A$ where $x: a\in D_{integer}$ and $y: a=5$, or as an $S$-formula $\forall x\forall y S_{a:=5;}(x,y)\Leftrightarrow x: a\in D_{integer}\land y:a=5$.

Let $x, y, y_1, y_2, \dots, y_n, z \in A$, where $A$ is abstract state set and let $a$ be a program variable of the type $Type$. The definitions of special $S$-relations no-operation, assignment, {\it if-then-else}, {\it if-then}, sequence and {\it while} are as follows:\\

\begin{definition} [No-operation]\label{def:no-operation}
$S$-relation $S_{nop}$ is defined as:
\begin{itemize}
\item[a.)] set $S_{nop}=\{(x,y) | x=y\}$, or
\item[b.)] $S$-formula $\forall x\forall yS_{nop}(x,y) \ \Leftrightarrow \ x=y$.
\end{itemize}
\end{definition}

\begin{definition} [Assignment]\label{def:assignment}
$S$-relation $S_{a:=e;}$ is defined as:
\begin{itemize}
\item[a.)] set $S_{a:=e;}=\{(x,y) | x: a\in D_{Type} \land y: a=e\}$, or
\item[b.)] $S$-formula $\forall x\forall yS_{a:=e;}(x,y) \ \Leftrightarrow \ x: a\in D_{Type} \land y: a=e$.
\end{itemize}
\end{definition}

\begin{definition}[If-then-else]\label{def:if-then-else}
$S$-relation $S_{if-then-else}$ is defined as:
\begin{itemize}
\item[a.)] set $S_{if-then-else}=\{(x,y) | (B(x)\land S_1(x,y))\lor(\lnot B(x)\land S_2(x,y))\}$, or
\item[b.)] $S$-formula $\forall x\forall yS_{if-then-else}(x,y) \ \Leftrightarrow \ (B(x)\land S_1(x,y))\lor(\lnot B(x)\land S_2(x,y))$.
\end{itemize}
\end{definition}

\begin{definition}[If-then]\label{def:if-then}
$S$-relation $S_{if-then}$ is defined as:
\begin{itemize}
\item[a.)] set $S_{if-then}=\{(x,y) | (B(x)\land S(x,y))\lor(\lnot B(x)\land S_{nop}(x,y))\}$, or
\item[b.)] $S$-formula $\forall x\forall yS_{if-then}(x,y) \ \Leftrightarrow \ (B(x)\land S(x,y))\lor(\lnot B(x)\land S_{nop}(x,y))$.
\end{itemize}
\end{definition}

\begin{definition}[Sequence]\label{def:sequence}
$S$-relation $S_{[S_1;S_2]}$ is defined as:
\begin{itemize}
\item[a.)] set $S_{[S_1;S_2]}=\{(x,y) | \exists z(S_1(x,z)\land S_2(z,y))\}$, or
\item[b.)] $S$-formula $\forall x\forall yS_{[S_1;S_2]}(x,y) \ \Leftrightarrow \ \exists z(S_1(x,z)\land S_2(z,y))$.
\end{itemize}
\end{definition}

\begin{definition}[While]\label{def:while}
$S$-relation $S_{while}$ is defined as:
\begin{itemize}
\item[a.)] set $S_{while}=\{(x,y) | \exists y_1, y_2, \dots, y_nB(x)\land S(x,y_1)\land B(y_1)\land S(y_1,y_2)\land B(y_2)\land S(y_2,y_3)\land \dots \land B(y_n)\land S(y_n,y)\land \lnot B(y)\}$, or
\item[b.)] $S$-formula $\forall x\forall yS_{while}(x,y) \ \Leftrightarrow \ \exists y_1, y_2, \dots, y_nB(x)\land S(x,y_1)\land B(y_1)\land S(y_1,y_2)\land B(y_2)\land S(y_2,y_3)\land \dots \land B(y_n)\land S(y_n,y)\land \lnot B(y)$.
\end{itemize}
\end{definition}

In the previous section, the theorems that represent the general laws of Hoare logic were proven. In this section, the special syntax units were analyzed using $S$-relations introduced by the appropriate definitions. In this way, we have developed a mechanism for proving the correctness of syntax units with respect to the specification given. Every syntax unit is treated as an interpretation of the appropriate $S$-relation, and the program specification is modeled as an ordered pair of $S$-predicates $(P,Q)$, where $P$ is a precondition and $Q$ is a postcondition. Apparently, proving syntax unit correctness conforms to proving the validity of the appropriate $S$-formula containing $S$-relation $S$ and $S$-predicates $P$ and $Q$. In addition to generality, one of the main advantages of this approach is simplicity, because the proof procedures do not require complicated mathematical apparatus. In order to prove program correctness and/or a new theorem it is sufficient to know first-order predicate calculus.

Various theories related to program semantics description and analysis that use an interpreted set of states do not deal with variable declarations. The reason is that they use the vector of program variables to describe states, so the problem arises of how to deal with state descriptions when some program variables are not (yet) defined. Such a problem does not exist in the $S$-calculus because it relies on the abstract state space. For defining a variable declaration, we use an appropriate $S$-relation defined by:

\begin{definition} [Declaration]\label{def:declaration}
$S$-relation $S_{a:Type;}$ is defined as:
\begin{itemize}
\item[a.)] set $S_{a:Type;}=\{(x,y) | y: a\in D_{Type}\}$, or
\item[b.)] $S$-formula $\forall x\forall yS_{a:Type;}(x,y) \ \Leftrightarrow \ y: a\in D_{Type}$.
\end{itemize}
\end{definition}

The possibility of modeling variable declaration is of considerable importance, especially for languages where declaration is treated as an ordinary statement (as in C/C++), and even may appear anywhere in the source code (as in Java). Using the appropriate $S$-relation enables us to automate verification of such programs. Let us consider simple two programs written in C.

\subsection{Example}\label{subsec:example1}

Is the syntax unit written in C and given in Figure \ref{fig:example1} correct with respect to the specification given as a pair of predicates $(\top,a=10)$?\\
\begin{figure}
\centerline{\parbox{104pt}{% it's safe to overestimate the size here
\begin{tabbing}
{\bf int} a=5;\\
{\bf if} (a $>$ 0)\\
\hspace*{1em} a=10;\\
{\bf else}\\
\hspace*{1em} a=100;
\end{tabbing}}}
\caption{Example \ref{subsec:example1}}
\label{fig:example1}
\end{figure}
The given syntax unit is an interpretation of $S$-relation $S$:\\ 

\noindent$S: \ a:integer; \ a:=5; \ if \ a>0 \ then \ a:=10 \ else \ a:=100;$\\

The specification is an ordered pair of $S$-predicates $(P,Q)$ where the precondition is $P:\top$ and the postcondition is $Q:a=10$. Now, we have to prove the validity of the $S$-formula $\{P\}S\{Q\} \ \leftrightarrow \ \forall x[P(x)\Rightarrow (\exists y S(x,y) \land \forall z(S(x,z) \Rightarrow Q(z)))] $. Apart from $P$, $S$ and $Q$, we will use the following notation:\\

\noindent$S_1: \ a:integer;$\\
$S_2: \ a:=5; \ if \ a>0 \ then \ a:=10 \ else \ a:=100;$\\
$S_3: \ a:=5;$\\
$S_4: \ if \ a>0 \ then \ a:=10 \ else \ a:=100;$\\
$S_5: \ a:=10;$\\
$S_6: \ a:=100;$\\
$R: \ a\in D_{integer}$\\
$T: \ a=5$\\
$B: \ a>0$\\
$W: \ a=10$\\
$U: \ a=100$\\

We will prove total correctness of the $S$-relation $S$ with respect to the specification $(\tau,W)$, i.e. we will prove that the formula $\{\tau\}S\{W\}$ is valid. The $S$-relation $S$ is a sequence $[S_{1};S_{2}]$. Consequently, according to the Definition \ref{def:sequence}, the following formula is valid:\\

\noindent$\forall x\forall yS(x,y)\Leftrightarrow \exists zS_1(x,z)\land S_2(z,y)$.\hfill$(1)$\\

Further, we prove that the $S$-relation $S_1$ is totally correct with respect to the specification $(\tau,R)$. Since $S_1$ is declaration of the program variable $a:integer$, according to the Definition \ref{def:declaration}, the following $S$-formulas are valid:\\

\noindent$\forall x\forall yS_1(x,y)\Leftrightarrow y: a\in D_{integer}$,\hfill$(2)$\\
$\{\tau\}S_1\{R\}$.\hfill$(3)$\\

We proceed by proving that the $S$-relation $S_2$ is correct with respect to the specification $(R,W)$, i.e. that the $S$-formula $\{R\}S\{W\}$ is valid. The $S$-relation $S_2$ is sequence $[S_{3};S_{4}]$. According to the Definition \ref{def:sequence}, it follows that the $S$-formula\\

\noindent$\forall x\forall yS_2(x,y)\Leftrightarrow\exists zS_3(x,z)\land S_4(z,y)$\hfill$(4)$\\

\noindent is valid. The $S$-relation $S_3$ is assignment. According to the Definition \ref{def:assignment}, the following formulas are valid:\\

\noindent$\forall x\forall zS_3(x,z)\Leftrightarrow x: a\in D_{integer}\land z: a=5$,\hfill$(5)$\\
$\{R\}S_3\{T\}$,\hfill$(6)$\\
$B(z)$.\hfill$(7)$\\

The $S$-relations $S_5$ and $S_6$ are assignments. From the Definition \ref{def:assignment}, it follows that the formulas $(8)$ - $(11)$ are also valid:\\

\noindent$\forall z\forall yS_5(z,y)\Leftrightarrow z: a\in D_{integer}\land y: a=10$,\hfill$(8)$\\
$\{R\}S_5\{W\}$,\hfill$(9)$\\
$\forall z\forall yS_6(z,y)\Leftrightarrow z: a\in D_{integer}\land y: a=100$,\hfill$(10)$\\
$\{R\}S_6\{U\}$.\hfill$(11)$\\

Since $T\Rightarrow R$, by the Theorem \ref{theorem:conseqence}$.a.)$, the following formulas are valid:\\

\noindent$\{T\}S_5\{W\}$,\hfill$(12)$\\
$\{T\}S_6\{U\}$.\hfill$(13)$\\

The $S$-relation $S_4$ is {\it if-then-else} so, by the Definition \ref{def:if-then-else}, the following formula is valid:\\

\noindent$\forall z\forall yS_4(z,y)\Leftrightarrow(B(z)\land S_5(z,y))\lor(\lnot B(z)\land S_6(z,y))$.\hfill$(14)$\\

From $(7)$, $(12)$ and $(14)$ we infer the validity of the formula:\\

\noindent$\{T\}S_4\{W\}$.\hfill$(15)$\\

From $(4)$, $(6)$ and $(15)$ we conclude:\\

\noindent$\{R\}S_2\{W\}$.\hfill$(16)$\\

From $(1)$, $(3)$ and $(16)$ we conclude:\\

\noindent$\{\tau\}S\{W\}$.\\

Since $P\Rightarrow\tau$ and $W\Rightarrow Q$, from the Theorem \ref{theorem:conseqence}$.c.)$ we conclude that the formula\\

\noindent$\{P\}S\{Q\}$\\

\noindent is valid, Q.E.D.\\

\subsection{Example}\label{subsec:example2}

Is the syntax unit written in C and given in Figure \ref{fig:example2} correct with respect to the specification given as a pair of predicates $(i=2\land n=4\land f=1, f=24)$?\\
\begin{figure}
\centerline{\parbox{104pt}{% it's safe to overestimate the size here
\begin{tabbing}
{\bf while} (i $<=$ n) \{\\
\hspace*{1em} f*=i;\\
\hspace*{1em} i++;\\
\}
\end{tabbing}}}
\caption{Example \ref{subsec:example2}}
\label{fig:example2}
\end{figure}
The given syntax unit is an interpretation of $S$-relation $S$:\\

$S: \ while \ i<=n \ do \ begin \ f:=f*i; \ i:=i+1; \ end;$\\

We have to prove the validity of $\{P\}S\{Q\}$, where $P: i=2\land n=4\land f=1$ and $Q: f=24$. Apart from $S$, $P$ and $Q$, we will use the following notation:\\

\noindent$R: \ i=5\land n=4\land f=24$\\
$B: \ i<=n$\\
$S_1: \ f:=f*i; \ i:=i+1;$\\
$S_2: \ f:=f*i;$\\
$S_3: \ i:=i+1;$\\

According to the Definition \ref{def:assignment}, the following formulas are valid:\\

\noindent$\forall x\forall z_1S_2(x,z_1) \ \Leftrightarrow \ x: i=2\land n=4\land f=1 \ \land \ z_1: i=2\land n=4\land f=2$,\\
$\forall z_1\forall y_1S_3(z_1,y_1) \ \Leftrightarrow \ z_1: i=2\land n=4\land f=2 \ \land \ y_1: i=3\land n=4\land f=2$\\

and by the Definition \ref{def:sequence}, we obtain:\\

\noindent$\forall x\forall y_1S_1(x,y_1) \ \Leftrightarrow \ x: i=2\land n=4\land f=1 \ \land \ y_1: i=3\land n=4\land f=2$.\hfill$(17)$\\

From the $S$-formula $(17)$ we conclude that the following formulas are valid:\\

\noindent$\forall xB(x)$,\hfill$(18)$\\
$\forall y_1B(y_1)$.\hfill$(19)$\\

According to the Definition \ref{def:assignment}, the following formulas are valid:\\

\noindent$\forall y_1\forall z_2S_2(y_1,z_2) \ \Leftrightarrow \ y_1: i=3\land n=4\land f=2 \ \land \ z_2: i=3\land n=4\land f=6$,\\
$\forall z_2\forall y_2S_3(z_2,y_2) \ \Leftrightarrow \ z_2: i=3\land n=4\land f=6 \ \land \ y_2: i=4\land n=4\land f=6$\\

and by the Definition \ref{def:sequence}, we obtain:\\

\noindent$\forall y_1\forall y_2S_1(y_1,y_2) \ \Leftrightarrow \ y_1: i=3\land n=4\land f=2 \ \land \ y_2: i=4\land n=4\land f=6$.\hfill$(20)$\\

From the $S$-formula $(20)$ we conclude that the following formula is valid:\\

\noindent$\forall y_2B(y_2)$.\hfill$(21)$\\

According to the Definition \ref{def:assignment}, the following formulas are valid:\\

\noindent$\forall y_2\forall z_3S_2(y_2,z_3) \ \Leftrightarrow \ y_2: i=4\land n=4\land f=6 \ \land \ z_3: i=4\land n=4\land f=24$,\\
$\forall z_3\forall yS_3(z_3,y) \ \Leftrightarrow \ z_3: i=4\land n=4\land f=24 \ \land \ y: i=5\land n=4\land f=24$\\

and by the Definition \ref{def:sequence}, we obtain:\\

\noindent$\forall y_2\forall yS_1(y_2,y) \ \Leftrightarrow \ y_2: i=4\land n=4\land f=6 \ \land \ y: i=5\land n=4\land f=24$.\hfill$(22)$\\

From the $S$-formula $(22)$ we conclude that the following formula is valid:\\

\noindent$\forall y\lnot B(y)$.\hfill$(23)$\\

From the $S$-formulas $(17)$ - $(23)$ we conclude that the following formula is valid:\\

\noindent$\forall x\forall yS(x,y) \ \Leftrightarrow \ \exists y_1, y_2B(x)\land S_1(x,y_1)\land B(y_1)\land S_1(y_1,y_2)\land B(y_2)\land S_1(y_2,y)\land \lnot B(y)$\\

and conclude that the following formula is valid:\\

\noindent$\{P\}S\{R\}$.\\

Since $R\Rightarrow Q$, from the Theorem \ref{theorem:conseqence}$.b.)$ we conclude that the formula\\

\noindent$\{P\}S\{Q\}$\\

\noindent is valid, Q.E.D.

\section{Proofs of Dijkstra's Theorems on the Weakest Precondition}\label{sec:fifth}

In the previous two sections, we have shown how to apply the $S$-calculus to prove program correctness. In this section, we will demonstrate the use of $S$-calculus for proving general theorems in a strictly formal way. As an example, we will consider Dijkstra's theorems on the weakest precondition $wp$, namely the laws of the excluded miracle, monotonicity, conjunction and disjunction. While being correct, the original proofs \cite{14}\cite{18} are not strictly formal, so our task will be to provide full formalization. In addition, we will establish yet another general law, the law of negation (Theorem \ref{theorem:wp_negation}), in order to provide a more complete insight to the behavior of the operator $wp$. The section ends with the formal proof of Dijkstra's theorem on total correctness. Some of the above-mentioned theorems will be proven in {\tt Coq}.

\begin{definition} [The Weakest Precondition]\label{def:wp}
The weakest precondition of $S$-relation $S$ with respect to postcondition $Q$ is $S$-predicate $wp(S,Q)$ if:\\
\noindent $(WP_1) \hspace{1cm} \{wp(S,Q)\}S\{Q\}$,\\
\noindent $(WP_2) \hspace{1cm} \{P\}S\{Q\} \ \Rightarrow \ \forall x(P(x)\Rightarrow wp(S,Q)(x))$.\\
\end{definition}

$ $\phantom{We prove theorems:}

\begin{theorem}[Dijkstra's Law of the Excluded Miracle]\label{theorem:wp_excluded_miracle}
The following $S$-formula is valid:

\vspace{0.215cm}\hspace{0.15cm} $wp(S,\phi)\Leftrightarrow\phi$.\\

{\bf Proof.}
$ $\phantom{We prove theorems:}
\begin{itemize}
\item[ ] According to $(WP_1)$ in the Definition \ref{def:wp}, we obtain:\\
$\{wp(S,\phi)\}S\{\phi\}$\\
and by the Theorem \ref{theorem:excluded_miracle}, we conclude that the following formula is valid:\\
$wp(S,\phi)\Leftrightarrow\phi$.\\
\end{itemize}
$\square$
\end{theorem}

\begin{theorem}[Dijkstra's Law of Monotonicity]\label{theorem:wp_monotonicity}
The following $S$-formula is valid:

\vspace{0.215cm}\hspace{0.15cm} $(Q\Rightarrow R) \ \Rightarrow \ (wp(S,Q)\Rightarrow wp(S,R))$.\\

{\bf Proof.}
$ $\phantom{We prove theorems:}
\begin{itemize}
\item[ ] According to $(WP_1)$ in the Definition \ref{def:wp}, we obtain:\\
$\{wp(S,Q)\}S\{Q\}$\\
and by the Theorem \ref{theorem:conseqence}$.b.)$ we conclude that the following formula is valid:\\
$\{wp(S,Q)\}S\{Q\}\land\forall x(Q(x)\Rightarrow R(x)) \ \Rightarrow \ \{wp(S,Q)\}S\{R\}$.\\
According to $(WP_2)$ in the Definition \ref{def:wp}, we obtain:\\
$(3) \ wp(S,Q)\Rightarrow wp(S,R)$.\\
\end{itemize}
$\square$
\end{theorem}

\begin{theorem}[Dijkstra's Law of Conjunction]\label{theorem:wp_conjunction}
The following $S$-formula is valid:

\vspace{0.215cm}\hspace{0.15cm} $wp(S,Q)\land wp(S,R) \ \Leftrightarrow \ wp(S,Q\land R)$.\\

{\bf Proof.}
$ $\phantom{We prove theorems:}
\begin{itemize}
\item[ ] First, let us prove the left-right implication:\\
$wp(S,Q)\land wp(S,R) \ \Rightarrow \ wp(S,Q\land R)$.\\
According to $(WP_1)$ in the Definition \ref{def:wp}, we obtain the following $S$-formulas:\\
$\{wp(S,Q)\}S\{Q\}$,\\
$\{wp(S,R)\}S\{R\}$.\\
By the Theorem \ref{theorem:conjunction}$.b.)$, we obtain:\\
$\{wp(S,Q)\}S\{Q\}\land\{wp(S,R)\}S\{R\} \ \Rightarrow \ \{wp(S,Q)\land wp(S,R)\}S\{Q\land R\}$\\
and according to $(WP_2)$ in the Definition \ref{def:wp}, we obtain:\\
$wp(S,Q)\land wp(S,R) \ \Rightarrow \ wp(S,Q\land R)$\\
and conclude that the left-right implication is valid.\\
Further, let us prove the right-left implication:\\
$wp(S,Q\land R) \ \Rightarrow \ wp(S,Q)\land wp(S,R)$.\\
According to $(WP_1)$ in the Definition \ref{def:wp}, we obtain following $S$-formula:\\
$\{wp(S,Q\land R)\}S\{Q\land R\}$.\\
By the Theorem \ref{theorem:conjunction_disjunction}$.b.)$, we obtain:\\
$\{wp(S,Q\land R)\}S\{Q\} \land \{wp(S,Q\land R)\}S\{R\}$\\
and according to $(WP_2)$ in the Definition \ref{def:wp}, we obtain:\\
$(wp(S,Q\land R)\Rightarrow wp(S,Q)) \ \land \ (wp(S,Q\land R)\Rightarrow wp(S,R))$.\\
After that, by the Theorem $(T_{19})$, we obtain:\\
$wp(S,Q\land R) \ \Rightarrow \ wp(S,Q)\land wp(S,R)$\\
and conclude that the right-left implication is valid.\\
Since both implications are valid, we conclude that the starting equivalence is valid.\\
\end{itemize}
$\square$\\
\end{theorem}

\begin{theorem}[Dijkstra's Law of Disjunction]\label{theorem:wp_disjunction}
The following $S$-formula is valid:

\vspace{0.215cm}\hspace{0.15cm} $wp(S,Q)\lor wp(S,R) \ \Rightarrow \ wp(S,Q\lor R)$.\\

{\bf Proof.}
$ $\phantom{We prove theorems:}
\begin{itemize}
\item[ ] By the Theorem \ref{theorem:conjunction}$.a.)$, we obtain:\\
$\{wp(S,Q)\}S\{Q\}\land\{wp(S,R)\}S\{R\} \ \Rightarrow \ \{wp(S,Q)\lor wp(S,R)\}S\{Q\lor R\}$\\
and according to $(WP_2)$ in the Definition \ref{def:wp}, we obtain:\\
$wp(S,Q)\lor wp(S,R) \ \Rightarrow \ wp(S,Q\lor R)$.\\
\end{itemize}
$\square$\\
\end{theorem}

\begin{theorem}[Law of Negation]\label{theorem:wp_negation}
The following $S$-formula is valid:

\vspace{0.215cm}\hspace{0.15cm} $\lnot(wp(S,Q)\land wp(S,\lnot Q))$.\\

{\bf Proof.}
$ $\phantom{We prove theorems:}
\begin{itemize}
\item[ ] According to $(WP_1)$ in the Definition \ref{def:wp}, if we substitute $P$ with $wp(S,Q)$ and $R$ with $wp(S,\lnot Q)$ in the Theorem \ref{theorem:negation}$.a.)$, we obtain:\\
$\lnot(wp(S,Q)\land wp(S,\lnot Q)$.
\end{itemize}
$\square$
\end{theorem}

\begin{theorem}[Dijkstra's Theorem on Total Correctness]\label{theorem:total_correctness}
The following $S$-formula is valid:

\vspace{0.215cm}\hspace{0.15cm} $\{P\}S\{Q\} \ \Leftrightarrow \ (P\Rightarrow wp(S,Q))$.\\

{\bf Proof.}
$ $\phantom{We prove theorems:}
\begin{itemize}
\item[ ] First, let us prove the left-right implication:\\
$\{P\}S\{Q\} \ \Rightarrow \ (P\Rightarrow wp(S,Q))$.\\
According to $(WP_2)$ in the Definition \ref{def:wp}, we conclude that the left-right implication is valid.\\
Further, let us prove the right-left implication:\\
$(P\Rightarrow wp(S,Q)) \ \Rightarrow \ \{P\}S\{Q\}$.\\
According to $(WP_1)$ in the Definition \ref{def:wp}, we obtain following $S$-formula:\\
$\{wp(S,Q)\}S\{Q\}$.\\
If we substitute $R$ with $wp(S,Q)$ in the Theorem \ref{theorem:conseqence}$.a.)$, we obtain:\\
$\forall x(P(x)\Rightarrow wp(S,Q)(x))\land \{wp(S,Q)\}S\{Q\} \ \Rightarrow \ \{P\}S\{Q\}$\\
and conclude that the right-left implication is valid.\\
Since both implications are valid, we conclude that the starting equivalence is valid.
\end{itemize}
$\square$
\end{theorem}

The proves of Theorems \ref{theorem:wp_excluded_miracle}, \ref{theorem:wp_negation} and \ref{theorem:total_correctness} in {\tt Coq} are given in Appendix \ref{appendix}.

\section{Conclusions}
In this paper, we have developed the $S$-calculus, which represents a powerful mathematical tool for program semantics analysis. The $S$-calculus is based on axioms and theorems of first-order predicate logic and uses $S$-formulas, which are defined on the abstract state space of a virtual machine. Proving program correctness and/or establishing new theorems conform to proving the validity of the appropriate $S$-formula, and for that, we need only the first-order predicate logic. Since the problem of indeterminism does not exist, $S$-calculus can consider total/partial correctness without the need for additional concepts and formulas. Owing to its generality, the $S$-calculus can cope with the semantics of variable declaration, where some other theories fail (the ones that are based on the interpreted set of states).

The mathematical mechanism developed in this paper, apart from being general, brings together Hoare's ideas and first-order predicate logic. It also enables automatic proofs of program correctness and/or new theorems. In this paper, we have provided strictly formal proofs for the general laws of Hoare logic and Dijkstra's theorems on the weakest precondition. Moreover, we have proven an additional law related to the weakest precondition, namely the law of negation. As an example, some of the theorems are proven using {\tt Coq} automatic prover.

Our future research in the area of $S$-calculus will be aimed towards investigating more complex properties and relationships that exist between preconditions, postconditions and syntax units, especially in the object-oriented environment. The second line of work will be development of algorithms for automated program correctness proofs, thus providing a practical aspect to $S$-calculus.

\newpage
\appendix \section{Appendix}\label{appendix}

{\rm \noindent Theorem \ref{theorem:conseqence}$.a.)$:}\\
{\tt $\phantom{dfgdfg}$}\\
{\tt Variable A: Set.}\\
{\tt Variables P Q R: A-$>$Prop.}\\
{\tt Variable S: A-$>$A-$>$Prop.}\\
{\tt Theorem t31a : \\ ((forall x:A,(P x-$>$R x)) $\slash\backslash$ \\ (forall x:A, (R x -$>$((exists y:A, S x y)$\slash\backslash$(forall z:A, (S x z -$>$Q z )))))) -$>$ \\ \hspace{1cm} (forall x, (P x -$>$((exists y, S x y)$\slash\backslash$(forall z, (S x z -$>$Q z ))))).}\\
{\tt firstorder.}\\
{\tt $\phantom{dfgdfg}$}\\

{\rm \noindent Theorem \ref{theorem:conseqence}$.b.)$:}\\
{\tt $\phantom{dfgdfg}$}\\
{\tt Variable A: Set.}\\
{\tt Variables P Q R: A-$>$Prop.}\\
{\tt Variable S: A-$>$A-$>$Prop.}\\
{\tt Theorem t31b : \\ ((forall x:A, (P x -$>$((exists y:A, S x y)$\slash\backslash$(forall z:A, (S x z -$>$Q z ))))) $\slash\backslash$ \\ (forall z:A,(Q z-$>$R z))) -$>$ \\ (forall x, (P x -$>$((exists y, S x y)$\slash\backslash$(forall z, (S x z -$>$R z ))))).}\\
{\tt firstorder.}\\
{\tt $\phantom{dfgdfg}$}\\

{\rm \noindent Theorem \ref{theorem:conseqence}$.c.)$:}\\
{\tt $\phantom{dfgdfg}$}\\
{\tt Variable A: Set.}\\
{\tt Variables P Q U V: A-$>$Prop.}\\
{\tt Variable S: A-$>$A-$>$Prop.}\\
{\tt Theorem t31c : \\ ((forall x:A,(U x-$>$P x)) $\slash\backslash$ \\ (forall x:A, (P x -$>$((exists y:A, S x y)$\slash\backslash$(forall z:A, (S x z -$>$Q z ))))) $\slash\backslash$ \\ (forall z:A,(Q z-$>$V z))) -$>$ \\ (forall x, (U x -$>$((exists y, S x y)$\slash\backslash$(forall z, (S x z -$>$V z ))))).}\\
{\tt firstorder.}\\
{\tt $\phantom{dfgdfg}$}\\

{\rm \noindent Theorem \ref{theorem:conjunction}$.a.)$:}\\
{\tt $\phantom{dfgdfg}$}\\
{\tt Variable A: Set.}\\
{\tt Variables P R Q W: A-$>$Prop.}\\
{\tt Variable S: A-$>$A-$>$Prop.}\\
{\tt Theorem t32a : \\ ((forall x:A, (P x -$>$ ((exists y:A, S x y)$\slash\backslash$(forall z:A, (S x z-$>$Q z ))))) $\slash\backslash$ \\ (forall x:A, (R x -$>$ ((exists y:A, S x y)$\slash\backslash$(forall z:A, (S x z-$>$W z )))))) -$>$ \\ (forall x:A, ((P x $\backslash\slash$ R x) -$>$ ((exists y:A, S x y)$\slash\backslash$(forall z:A, (S x z-$>$(Q z $\backslash\slash$ W z)))))).\\ 
{\tt firstorder.}\\
{\tt $\phantom{dfgdfg}$}\\

{\rm \noindent Theorem \ref{theorem:conjunction}$.b.)$:}\\
{\tt $\phantom{dfgdfg}$}\\
{\tt Variable A: Set.}\\
{\tt Variables P Q R W: A-$>$Prop.}\\
{\tt Variable S: A-$>$A-$>$Prop.}\\
{\tt Theorem t32b : \\ ((forall x:A, (P x -$>$ ((exists y:A, S x y)$\slash\backslash$(forall z:A, (S x z-$>$Q z ))))) $\slash\backslash$ \\ (forall x:A, (R x -$>$ ((exists y:A, S x y)$\slash\backslash$(forall z:A, (S x z-$>$W z )))))) -$>$ \\ (forall x:A, ((P x $\slash\backslash$ R x) -$>$ ((exists y:A, S x y)$\slash\backslash$(forall z:A, (S x z-$>$(Q z $\slash\backslash$ W z)))))).}\\ 
{\tt firstorder.}\\
{\tt $\phantom{dfgdfg}$}\\

{\rm \noindent Theorem \ref{theorem:disjunction}:}\\
{\tt $\phantom{dfgdfg}$}\\
{\tt Variable A: Set.}\\
{\tt Variables P Q R W: A-$>$Prop.}\\
{\tt Variable S: A-$>$A-$>$Prop.}\\
{\tt Theorem t33 : \\ ((forall x:A, (P x -$>$ ((exists y:A, S x y)$\slash\backslash$(forall z:A, (S x z-$>$Q z ))))) $\backslash\slash$ \\ (forall x:A, (R x -$>$ ((exists y:A, S x y)$\slash\backslash$(forall z:A, (S x z-$>$W z )))))) -$>$ \\ (forall x:A, ((P x $\slash\backslash$ R x) -$>$ ((exists y:A, S x y)$\slash\backslash$(forall z:A, (S x z-$>$(Q z $\backslash$\slash W z)))))).}\\
{\tt firstorder.}\\
{\tt $\phantom{dfgdfg}$}\\

{\rm \noindent Theorem \ref{theorem:excluded_miracle}:}\\
{\tt $\phantom{dfgdfg}$}\\
{\tt Variable A: Set.}\\
{\tt Variables P : A-$>$Prop.}\\
{\tt Variable S: A-$>$A-$>$Prop.}\\
{\tt Definition phi (x:A) := False.}\\
{\tt Theorem t35 : \\ (forall x:A, (P x -$>$ ((exists y:A, S x y)$\slash\backslash$(forall z:A, (S x z-$>$phi z)))) $<$-$>$  \\ (P x$<$-$>$phi x)).}\\
{\tt firstorder.}\\
{\tt $\phantom{dfgdfg}$}\\

{\rm \noindent Theorem \ref{theorem:negation}$.a)$:}\\
{\tt $\phantom{dfgdfg}$}\\
{\tt Variable A: Set.}\\
{\tt Variables P Q R: A-$>$Prop.}\\
{\tt Variable S: A-$>$A-$>$Prop.}\\
{\tt Theorem t36a : \\ ((forall x:A, (P x -$>$ ((exists y:A, S x y)$\slash\backslash$(forall z:A, (S x z-$>$Q z))))) $\slash\backslash$  \\ (forall x:A, (R x -$>$ ((exists y:A, S x y)$\slash\backslash$(forall z:A, (S x z-$>$($\sim$Q z))))))) -$>$  \\ (forall x:A, ($\sim$(P x $\slash\backslash$ R x))).}\\ 
{\tt firstorder.}\\
{\tt $\phantom{dfgdfg}$}\\
{\tt $\phantom{dfgdfg}$}\\
{\tt $\phantom{dfgdfg}$}\\

{\rm \noindent Theorem \ref{theorem:negation}$.b)$:}\\
{\tt $\phantom{dfgdfg}$}\\
{\tt Variable A: Set.}\\
{\tt Variables P Q: A-$>$Prop.}\\
{\tt Variable S: A-$>$A-$>$Prop.}\\
{\tt Theorem t36b : \\ ((forall x:A, (P x -$>$ ((exists y:A, S x y)$\slash\backslash$(forall z:A, (S x z-$>$Q z))))) $\slash\backslash$ \\ (forall x:A, (P x -$>$ ((exists y:A, S x y)$\slash\backslash$(forall z:A, (S x z-$>$($\sim$Q z))))))) $<$-$>$ \\ (forall x:A, ($\sim$(P x))).}\\
{\tt firstorder.}\\
{\tt $\phantom{dfgdfg}$}\\
{\tt $\phantom{dfgdfg}$}\\
{\tt $\phantom{dfgdfg}$}\\

{\rm \noindent Theorem \ref{theorem:wp_excluded_miracle}:}\\
{\tt $\phantom{dfgdfg}$}\\
{\tt Variable A: Set.}\\
{\tt Variables P wpSphi: A-$>$Prop.}\\
{\tt Variable S: A-$>$A-$>$Prop.}\\
{\tt Definition phi (x:A) := False.}\\
{\tt Axiom wpSphi1 : \\ forall x:A, (wpSphi x -$>$ ((exists y:A, S x y)$\slash\backslash$(forall z:A, (S x z-$>$phi z)))).}\\
{\tt Axiom wpSphi2 : \\ forall x:A, (P x -$>$ ((exists y:A, S x y)$\slash\backslash$(forall z:A, (S x z-$>$phi z)))) -$>$ \\ (P x-$>$wpSphi x).}\\
{\tt Theorem t51 : forall x:A, (wpSphi x $<$-$>$ phi x).}\\
{\tt firstorder using wpSphi1 wpSphi2.}\\
{\tt $\phantom{dfgdfg}$}\\
{\tt $\phantom{dfgdfg}$}\\
{\tt $\phantom{dfgdfg}$}\\

{\rm \noindent Theorem \ref{theorem:wp_negation}:}\\
{\tt $\phantom{dfgdfg}$}\\
{\tt Variable A: Set.}\\
{\tt Variables P Q R wpSQ wpSNQ: A-$>$Prop.}\\
{\tt Variable S: A-$>$A-$>$Prop.}\\
{\tt Axiom wpSQ1 : \\ forall x:A, (wpSQ x -$>$ ((exists y:A, S x y)$\slash\backslash$(forall z:A, (S x z-$>$Q z)))).}\\
{\tt Axiom wpSQ2 : \\ forall x:A, (P x -$>$ ((exists y:A, S x y)$\slash\backslash$(forall z:A, (S x z-$>$Q z)))) -$>$ \\ (P x-$>$wpSQ x).}\\
{\tt Axiom wpSNQ1 : \\ forall x:A, (wpSNQ x -$>$ ((exists y:A, S x y)$\slash\backslash$(forall z:A, (S x z-$>$($\sim$(Q z)))))).}\\
{\tt Axiom wpSNQ2 : \\ forall x:A, (P x -$>$ ((exists y:A, S x y)$\slash\backslash$(forall z:A, (S x z-$>$($\sim$(Q z)))))) -$>$ \\ (P x-$>$wpSNQ x).}\\
{\tt Theorem t55 : forall x:A, ($\sim$((wpSQ x)$\slash\backslash$(wpSNQ x))).}\\ 
{\tt firstorder using wpSQ1 wpSQ2 wpSNQ1 wpSNQ2.}\\
{\tt $\phantom{dfgdfg}$}\\
{\tt $\phantom{dfgdfg}$}\\
{\tt $\phantom{dfgdfg}$}\\

{\rm \noindent Theorem \ref{theorem:total_correctness}:}\\
{\tt $\phantom{dfgdfg}$}\\
{\tt Variable A: Set.}\\
{\tt Variables P Q wpSQ: A-$>$Prop.}\\
{\tt Variable S: A-$>$A-$>$Prop.}\\
{\tt Axiom wpSQ1 : \\ forall x:A, (wpSQ x -$>$ ((exists y:A, S x y)$\slash\backslash$(forall z:A, (S x z-$>$Q z)))).}\\
{\tt Axiom wpSQ2 : \\ forall x:A, (P x -$>$ ((exists y:A, S x y)$\slash\backslash$(forall z:A, (S x z-$>$Q z)))) -$>$ \\ (P x-$>$wpSQ x).}\\
{\tt Theorem t56 : \\ forall x:A, ((P x -$>$ ((exists y:A, S x y)$\slash\backslash$(forall z:A, (S x z-$>$Q z)))) $<$-$>$ \\ ((P x)-$>$(wpSQ x))).}\\
{\tt firstorder using wpSQ1 wpSQ2.}

\end{document}